\documentclass{article}

\usepackage{amsmath}

\usepackage{amscd}

\usepackage{amsthm}

\usepackage{amssymb} \usepackage{latexsym}

\usepackage{eufrak}

\usepackage{euscript}

\usepackage{psfig}

\usepackage{epsfig}

\usepackage{graphics}

\usepackage{array}

\usepackage{enumerate}

\newcommand{\bel}[1]{\begin{equation}\label{#1}}

\newcommand{\be}{\begin{equation}}

\newcommand{\ee}{\end{equation}}

\newcommand{\ba}{\begin{eqnarray}}

\newcommand{\ea}{\end{eqnarray}}

\newcommand{\qe}{\end{equation}}

\newcommand{\q}{equilibrium}

\begin{document}

\newtheorem{theorem}{Theorem}

\title{
Individual strategies in complementarity games and population
dynamics
 \\
\vspace{2.5cm}
}

\author{J\"urgen Jost\footnote{Max Planck Institute for Mathematics in the
  Sciences, Inselstr.22-26, 04103 Leipzig, Germany, jost@mis.mpg.de}
\footnote{Santa Fe Institute, 1399 Hyde Park Road, Santa Fe, NM
87501, USA, jost@santafe.edu}\\
Wei Li\footnote{Max Planck Institute for Mathematics in the
Sciences, Inselstr.22-26, 04103 Leipzig, Germany,
liwei@mis.mpg.de} \footnote{Supported by the
Alexander-von-Humboldt Foundation} \footnote{On leave from
Hua-Zhong Normal University, Wuhan 430079, P.R. China}} \maketitle

\begin{abstract}
We introduce and study an evolutionary complementarity game where
in each round a player of population 1 is paired with a member of
population 2. The game is symmetric, and each player tries to
obtain an advantageous deal, but when one of them pushes too hard,
no deal at all can be concluded, and they both loose. The game has
many equilibria, and which of them is reached depends on the
history of the interactions as the players evolve according to a
fitness function that measures their gains across the
interactions. We can then break the symmetry by assigning
different strategy spaces to the populations, varying in
particular with respect to the information available to the
agents. The agents can, for example, adapt to the behavior of
their opponents met in previous rounds, or they can try to copy
the strategies of their successful friends. It turns out that, in
general, the more restricted strategy spaces, that is, those that
utilize less information, are more advantageous for a population
as a whole as their adoption drives the equilibrium in a direction
advantageous to that population. One reason is that a simpler
strategy can be learned faster in an evolutionary setting, another
is that it is good for a population to have some individuals that
are unfit in the sense that they make offers that are individually
unsuccessful, but have a systematic effect on the strategies of
their opponents. All these effects are demonstrated through
systematic simulations.
\end{abstract}


\section{Introduction}

We introduce a simple new game between members of two populations
that possesses some features hitherto not considered in game
theory and therefore exhibits some new aspects and offers
possibilities to
investigate new issues.\\
The game is simple: We have two populations, called the ``buyers''
and the ``sellers''. For each interaction, a member of one of them
is paired with a randomly drawn member of the other one. The buyer
offers a certain amount $x$ between 0 and $K$, and the seller asks
an amount $y$ within that same range. When $x \ge y$, a deal is
concluded where the buyer pays $x$, or, equivalently, gains $K-x$,
and the seller gains $y$. When $x < y$, they both gain nothing
(or, in the first
version, the buyer looses the amount $K$).\\
The rules are strict: In each interaction, the players do not know
what the opponent is offering or asking, and the own bid cannot be
adjusted. However, they can learn from experience. That is, the
players are allowed to develop strategies that consist in choosing
their bid in the present round for example on the basis of the bid
of the opponent met in the previous interaction. \\
The outcomes of these interactions then are  used to evaluate the
fitness of the players in an evolutionary process. After each
player has been chosen to play a given number of such
interactions, the fitness values, that is, the accumulated gains
of all the players in each population are compared,  and a
selection process is applied. That means that the chance for a
player of the present generation to be represented in the next
generation, or to be chosen as a parent in a recombination process
from a standard genetic algorithm, is proportional to his relative
fitness. In addition, we allow for mutations to occur with a
certain probability when building the next generation.\\
Of course, there is ample room for changing the rules of the game,
but let us analyze some important features of our game so that we
can see what we can learn from it.
\begin{enumerate}
\item The competition takes place at two different levels. Each
player plays against the members of the other population and in
order to avoid the risk that the interaction becomes a complete
failure, he has to make cautious bids. On the other hand, if a
bolder bid is successful, he achieves a higher fitness gain than
his more cautious fellows from his own population and thereby
gains a selective advantage over them. Putting it the other way
around, a higher gain in the case of a successful interaction is
offset by a higher risk for the failure of that interaction.
 \item Any value $k$ between 0 and $K$ can be an equilibrium in the
  following sense: When all the other members of the two populations
  always bid $k$, then any player choosing a bid value different from
  $k$ will be at a disadvantage: When he is a buyer and offers more
  than $k$, then he will pay more than the other buyers and become
  correspondingly less fit. When he offers less, then he cannot strike
  any deal and loses even more, namely the full amount $K$. For a
  seller, the situation is symmetric.
\item Whether an equilibrium is achieved, and what its value then
is, depends entirely on the history. When such an equilibrium
occurs at a value larger than $K/2$, the sellers are better off,
else the buyers.
 \item The situation between the populations is
symmetric, but we can then break the symmetry in various ways to
investigate the effect of certain parameters or strategic options.
For example, we could make the population sizes or the mutation
rates different. We could also allow one population to choose from
one class of possible strategies, and the other one from another
class. We can then see whether an equilibrium will be reached and
what its value is.
 \item Combining the insights from the previous points, in principle,
  if that were allowed in the game, one of the populations could
  coordinate its actions so as to drive the equilibrium in a direction
  favorable to it. For example, at an equilibrium at value $k$, the
  buyers could consistently offer an amount less than $k$. Eventually,
  the sellers would then react and also lower their bid
  correspondingly, to avoid the failure of all interactions. Now, we
  are not allowing such coordinated actions in our game, but also
  different mutation rates or different strategic options for the
  members of the two populations could drive the equilibrium value in
  a direction more favorable to one of the populations.
\end{enumerate}
As set up, our game is not a zero-sum game; it could be made into
one by letting the buyer pay, and the seller receive, the amount
$\frac{x+y}{2}$ in each successful interaction. Such a
modification should not substantially affect the above points. The
rule adopted by us could be interpreted as giving the difference
$x-y$ to some middleman or trader. One could also look at the
formally equivalent game where two partners with complementary
ingredients give the amounts $x$ and $z=K-y$ towards some common
cause; that cause succeeds only when $x+z \ge K$; in that case,
they each pay what they commit, that is $x$ or $z$, while when it
fails, they both suffer a loss of $K$. Again, they both wish to
minimize their costs or losses. Thus, with respect to those
individual interactions, our game is similar to, and in fact even
simpler than, standard cooperation games like the prisoner's
dilemma. One difference is that the number of options is much
larger here; the players can choose any number $k$ between 0 and
$K$. Even though in our simulations, we restrict $k$ to be an
integer, this does not cause much of a difference as $K$ can be
chosen rather large. The more important difference, and what makes
our set-up really interesting, is the division into two
complementary populations from which the players are taken. This
allows us to investigate how any symmetry breaking, for example
when one of the populations starts exploring a new set of
strategies, affects the resulting equilibrium, or perhaps even
prevents an equilibrium.  \\Thus, our game can serve to study the
efficiency of different strategy sets available to, or adopted by,
a population. The simplest meaningful type of strategy would be
for each player to simply select one fixed value $k$ and play that
value all the time. The selection process should then assure that
the two populations reach an equilibrium value. When random
fluctuations like mutations occur, one should expect some safety
margin, that is the buyers will settle on a slightly higher value
than the sellers. If we also allow learning from experience, we
naturally arrive at the following type of strategy: Each player
develops a look-up table from which he selects his current bid as
a function of the bid his opponent(s) in the previous round(s)
chose. Of course, this can get unwieldy if the players can
remember  too many rounds from the past. Here, however, already
the first non-trivial finding from our simulations becomes
relevant. Namely, we observe that a population whose members
remember only the value of the opponent's bid from the one most
recent round achieves a favorable \q\ value against a population
that can look two rounds back. There are two simple and apparently
general conceivalbe  reasons for this that are partly in conflict
with each other, but both of which seem to play a role here:
Firstly, when the set of available options is smaller, an optimal
strategy within that set can be found more quickly. In other
words, a more complicated strategy may have so many entries to
adapt that it takes such a long time to test them through exposure
in the game and subsequent fitness evaluation that by that time,
the other population with the simpler strategy space has already
settled into an advantageous state from which it cannot be
disposed anymore. Thus, a strategy that can be learned more
quickly beats another one that takes longer to develop, because,
according to the rules of our game, it can determine a favorable
equilibrium from which it cannot be driven anymore once attained.
Secondly, an explanation in the opposite direction is that with a
richer strategy set, a population should also be able to react
more accurately or quickly to fluctuations in the behavior of the
other population, and then, for example, the sellers could adjust
their own bids downwards more readily in response to some erratic
behavior of some of the buyers. This is similar to the red king
effect coined by Lachmann and Bergstrom\cite{BL} where it also is
advantageous to respond more slowly to the opponent's actions.
This finding may also have some interest in the context of the
rationality issue in economic theory. There, the ideal is an actor
that is in full command of all the available information when
selecting his actions. Of course, it is questionable to what
extent such an assumption is realistic, and this then becomes an
empirical question at the intersection of economics and
psychology. Our findings, however, shed a little light on this
issue from a different perspective. Namely, we observe that a
population whose individuals consistently adopt a less rational
strategy, in the sense that they utilize a smaller amount of the
available information from their own experience can do better in
competition with one whose members make more use of that
information, and so are more rational in the sense of economic
theory. Of course, the issue here is not that the individual agent
is doing worse when he tries to act in a cleverer manner, and in
fact, such an agent could certainly gain an individual advantage
in comparison with his more stupid fellows.  The point is rather
that a population of individuals that individually act more
stupidly can drive an \q\ between populations to its advantage.
\\Another choice of strategies is the more indirect one of imitating
successful fellows instead of directly responding to the bid of
the previous opponent. Formally, we introduce a network structure
into a population. In order to be able to compare different
network types, we fix a value $m$ for the average connectivity of
the nodes in the network. Those nodes represent the individual
members of the population, and an edge between two nodes stands
for some form of acquaintance or information sharing which we call
``friendship'' for simplicity. The friendship strategy then simply
consists in taking the average (or some weighted average in a more
refined version) of the bids of those friends of the player that
have had a successful interaction in the previous round. This
could, but need not, include the own previous bid in case it was
successful as well. If none of those bids from the previous round
has been successful, the player makes a random bid. This is, of
course, the standard option in all such situations where the
adopted strategy does not apply to compute a
bid. \\
We can then let different network topologies, like regular,
random, small-world or scale-free, play against each other. So
far, no network type has been consistently superior to all other
ones. When we let a population with such a friendship network
strategy play against one that chooses the one-round opponent bid
strategy, the latter in most cases, but not always, gets an
advantage, in the usual sense of an \q\ value that is more
favorable to it. Letting a friendship network play against a
two-round opponent strategy did not show a decisive
advantage on average for either of them. \\
Another issue that might play some role here is the homogeneity or
heterogeneity of a population. Since we are not allowing direct
coordination of the actions of the members of our population,
except indirectly as for example through a friendship network,
homogeneity of a population can indirectly enforce strong
similarities between the individual behaviors. This might then be
advantageous for driving the \q\ towards a value that is favorable
for the population in question. On the other hand, any such \q\ is
stable once it is reached and, in particular, predictability
cannot be exploited here. Therefore, the only chance to modify the
\q\ value after the transient period consists in triggering a
reaction of the opposite population through random or systematic
deviations in the preferred direction. This could be achieved
through mutations, or by other means for making the population
heterogeneous enough to always maintain some deviating strategies
by some of its members. Again, those deviating members will
probably be less successful than the more conforming ones, and so,
they will be eliminated by selection. The conclusion is that a
population as a whole can gain when it can develop a mechanism for
producing such individually less successful members. This leads to
the issue of group selection, amply discussed in theoretical
evolutionary biology, which, however, we do not intend to enter
here more deeply.

\section{Games between populations}

To clarify our setting, we discuss here how it fits into the
standard theory of games between two populations. A good reference
for that topic is \cite{VR}. There, such games are treated as
extensions of games within a single population. Players are
randomly paired, but each player faces an opponent from the
opposite population. It is instructive to subject the strategy
evolution to replicator dynamics, as systematically developed in
\cite{HS} and applied to 2-population games in \cite{VR}. To see
how this applies to our example, we start with the simplified
situation where each player has only two strategic options; to
achieve as much symmetry as possible, we play the complementarity
game described in the introduction as an equivalent version of our
buyer-seller game.\footnote{After interchanging the
  rows in the pay-off matrix introduced below, the game becomes
  equivalent to the trading complementarities game analyzed in 10.4.3.1
  of \cite{VR} who in turn refers to \cite{Di}. Thus, in particular,
  the analysis presented there applies to the problem at hand when
  played within a single population, and the extension to two
  populations is straightforward. The setting we adopt here, however,
  is better suited for our purpose of enlarging the strategy space.}
Thus, each player has the options to offer either 1 or 2, and when
the sum of the two offers is at least 3, each player receives 3
minus its own offer; else neither receives anything. Thus, the
pay-off matrix is

\[
\left (
\begin{array}{cc}
0 & 2\\
1 & 1
\end{array}
\right )
\]

Here, the $i$th row stands for the strategy $i$ of a player, the
$j$th column for the corresponding strategy of his opponent. When
$x^i$ denotes the proportion of players playing $i$ in the first
population, $y^i$ the corresponding number in the second one, we
can set up the corresponding replicator dynamics as
\begin{eqnarray}
\dot x^1 &=& x^1(2y^2 - x^1(2y^2) -x^2(y^1+y^2))\\
\dot x^2 &=& 1-\dot x^1
\end{eqnarray}

since $x^1+x^2=1$, and the symmetric formulae for the $y^j$.
Putting $x:=x^1$, $y:=y^1$ for simplicity, we arrive at the system

\begin{eqnarray}
\dot x &=&x (1-x)(1-2y)\\
\dot y &=&y(1-y)(1-2x)
\end{eqnarray}

This system has the two stable equilibria $(x,y)=(1,0)$ and
$(0,1)$, with the line $x=y$ separating the two basins of
attraction. Thus, whichever population has initially the larger
number of players of
strategy 1 will turn the game to its advantage. \\
The two basins of attraction will remain of equal size even when
one of the population adapts more slowly than the other one. In
the extreme case when the second one remains static, that is,
$\dot y=0$, the two basins of attraction\footnote{In this limiting
case, the attractors are no longer points, but line segments, but
that does not affect the discussion.} are separated by the line
$y=1/2$; when $y$ is below that value, $x$ will tend to 1, else to 0.\\
When we extend the game to allow three values for the offers, 1,2,
and 3, with a sum of 4 needed to give each player a pay-off of 4
minus its offer, the pay-off matrix is

\[
\left (
\begin{array}{ccc}
0 & 0 & 3\\
0 & 2 & 2\\
1 & 1 &1
\end{array}
\right )
\]

and the replicator dynamics for the strategies of the first
population is

\begin{eqnarray}
\dot x^1 &=& x^1(3y^3 - x^1(3y^3) -x^2(2y^2+2y^3)-x^3)\\
\dot x^2 &=& x^2(2y^2+2y^3 - x^1(3y^3) -x^2(2y^2+2y^3)-x^3)\\
\dot x^3 &=& x^3(1 - x^1(3y^3) -x^2(2y^2+2y^3)-x^3)
\end{eqnarray}

where we have used $y^1+y^2+y^3=1$ to simplify the coefficient of
$x^3$, and again symmetric formulae for the strategies $y^j$ of
the second population. So, that population whose  pay-off is
largest will also grow fastest. In particular, when all three
strategies are initially equally represented in the second
population, that is $y^1=y^2=y^3$, then $x^2$ will grow at the
fastest rate. Stable equilibria are at $x^1=1,y^3=1$,
$x^2=1,y^2=1$ and $x^3=1,y^1=1$, among which the middle one has
the largest basin of attraction. The reason is that it has the
highest pay-off when averaged over the strategies of the opposite
population. The pattern and its extension to more strategies is
obvious.
\\The following heuristic consideration is useful for
understanding some of the sequel better. As long as $y^3$ is not
the dominant strategy in the other population, $x^3$ has a higher
growth rate than $x^1$. Namely,

\be {(\frac{x^3}{x^1})}\dot \    = \frac{x^3}{x^1} (y^1 + y^2
-2y^3). \ee

Thus, for example when we are near the equilibrium point
$x^2=1,y^2=1$ and the second population is subjected to some
random perturbation while the first one then is allowed to adapt,
we expect a preference for strategy 3 over 1 to develop. Thus, the
population that can adapt more quickly will tend to prefer the
more cautious strategy 3 over the bolder strategy 1. The slower
population will then see an increase of 3 in the other population
and can thus increase the rate for its own strategy 1. Thus, the
situation will develop in its favor. This is similar to the red
king effect of \cite{BL}, with the difference that here not the
faster responding player is put at an individual disadvantage, but
rather the faster and individually beneficial responses of the
members of a population ultimately are disadvantageous to the
population as a whole and thus also for its
individual members.\\
Another reason why a faster evolving population could prefer
strategy 3 over 1 is a finite size effect. Namely, the players of
1 undergo a higher risk of being eliminated when their performance
over finitely many interactions is counted as their fitness.\\
However, the slight preference for 3 over 1 should not be seen as
a result of stochastic instability because  the stable and
dominant strategy is 2.\\
In any case, we should point out that neither of the preceding is
the correct explanation for the phenomena we see below where a
population with simpler strategies beats those with more complex
ones. Essentially, a more restricted set of strategic options will
focus the agents better in the relevant subset of the strategy
space. \\
In any case, if one population would have a central coordination
mechanism and sufficient foresight, it could clearly drive the
other, uncoordinated one into an equilibrium that is advantageous
for itself. However, we shall not allow any such thing in our
simulations.\\

\section{Metastrategies}

Thus, the standard replicator analysis of our game is rather
trivial. It depends on the initial relative frequencies of the
strategies in the two populations which of the many possible
equilibria is reached. As explained in the introduction, we wish
to apply here a different evolutionary setting than the one
encoded in the replicator dynamics. Namely, we offer the players
richer strategic options; for example, they could take their past
experiences into account, or follow their successful friends.
(Those particular options are similar to, but not identical with
those considered in \cite{VR}, called ``best response'' and
``imitation''.) The question addressed in our simulations then is
which kind of metastrategy when applied by all players in a
population will bring that population into the basin of
a favorable attractor.\\
One emergent feature of our simulation results is that typically a
population with a more restricted set of strategic options does
better in competition with one with a larger set. A reason for
this is that when there are fewer response options available each
single response has a higher frequency of being chosen and thereby
subjected to an evolutionary test. It thus has more opportunities
to exhibit its fitness and to adapt.\\
Let us consider the more concrete case where the members of one
population, labelled 1,  can only use the simplest possible
strategy, namely to choose one fixed value for their offer whereas
the members of the other population, called 2, are able to
determine their current bid on the basis of the bid of the
opponent from the previous encounter. That is, they have a look-up
vector with $K$ entries where the $k$th element gives their
response, that is, their current bid, when the previous opponent
had been asking $k$. Of course, in principle, such an agent could
just choose a constant vector, that is, select its current bid
independently of its previous experience, and so, the richer
strategy set includes the simpler one. Nevertheless, it turns out
that the resulting equilibrium is in favor of the population with
the simpler strategies. This can be understood as follows. The
members of 2 have only one value which then is subjected to many
evolutionary test, and so, the population 2 might rather quickly
settle their values within some restricted range. Population 1
will then only adapt those values of the look-up vectors that
correspond to those values exhibited by 2. A mutant from 2 with a
more risky strategy then, while having a lower expectation value
for its fitness, still has a positive probability to gain a higher
fitness because he can explore a region in which the members of 1
have not yet adapted and behave more or less randomly. Some more
members of population 2 can then follow when we renew the
population evolutionarily according to their fitness. This then
forces the members of 1 to adapt accordingly, because those that
do not bow to the more aggressive behavior of their opponents risk
being eliminated by our evolution algorithms since many failed
trades lower their fitness. The point is that the mutant from 2
were essentially facing a random behavior from 1 whereas the
members of 2 now are confronted with a more systematic behavior
from 2 which forces them to yield.

\section{Simulation Results}

In our simulations, we choose two populations of $N=400$
individuals each, the buyers and the sellers. In each round, each
buyer is randomly paired with a seller, and they both make offers
between 0 and $K=49$ or $K=99$ (only in one simulation $K$ is
chosen to be 19). After 1000 rounds, the population for the next
generation is determined, by letting the present population
reproduce differentially according to the fitness accumulated
during those rounds. When the strategy consists of a look-up
vector or matrix where the $i$th entry, or the $ij$ matrix element
tells the agent what to bid when his opponent from the previous
round bid $i$, or the two previous opponents bid $i$ and $j$,
resp., we also adopt genetic recombination as in standard genetic
algorithms to construct the next generation. We use a cross-over
probability of
0.70.\\
Before presenting the simulations of our game, we would like to
define some variables that will be used later. The $success$
$rate$ of deals taking place at time step $t$ is defined as $m(t)/
N$, where $m (t)$ represents the number of successful deals, that is,
those where the buyer offered at least as much as the seller asked, at
time $t$ and $N$ is the number of buyers or sellers. We define the
$median$ $fee$ as
 \be \frac
{\sum_{i=1}^{m (t)} Off_{buy} (i,t)} {m (t)} - \frac
{\sum_{i=1}^{m (t)} Off_{sel} (i,t)} {m (t)}, \ee
 where
$Off_{buy} (i,t) (Off_{sel} (i,t))$ is the offer of the $i$th
successful buyer (seller) at $t$. In addition, we use
$evolutionary$ $difference$ to monitor only the decreasing values
of $median$ $fee$, that is, its value at $t$ is the minimum of all
values of $median$ $fee$ between time steps 0 and $t$.\\
We start with the simplest conceivable strategy, namely the one
where each agent can only choose one fixed value for all his bids.
This value then is subjected to the standard evolutionary scheme,
that is, after a fixed number of rounds has been played, agents
reproduce
differentially according to their fitness.\\
So far, the situation is symmetric between the two populations,
but we can easily break the symmetry by introducing mutations
between generations and to assign different mutation rates to the
two populations. Thereby, we can understand the role of the random
mutation rate  in our game and find out  whether an optimized
value of this mutation rate exists and what its value then is.  In
most simulations, the population with a small random mutation
rate, like 0.01, is doing slightly worse than the opponent side
with a much higher mutation rate like 0.04 or 0.05 (see Fig. 1
(a)). We also observed that the side being subject to random
mutations with a very high rate, like 0.1, can do slightly worse
than the other side who is subject to  random mutations with
relatively lower rates, like 0.05, as in Fig. 1 (b). However, 0.04
or 0.05 is not the best value in all simulations, and there is no
exception-free conclusion possible here. Nevertheless, the values
from 0.01 to 0.05 will be good parameters, given the
population size of our models.\\
After these preparations, we can discuss  simulations with richer
strategic options for the players. First we come to the
simulations of our game where all the players, both the buyers and
the sellers, choose their bids on the basis of their own past
experiences. More specifically, players' offers at time $t+1$ are
functions of offers of their opponents met at time $t$. Say, if a
buyer's last-round opponent asks $k$, then his bid at the current
round will be the $k$th entry in the vector of his own look-up
table. In these simulations, the random mutation rate is 0.01.
Fig. 2 (a) shows in one simulation the variation of success rate
with respect to the time $t$. As we can see from the figure, the
success rate is improved very quickly. After only 15 generations,
the success rate has been improved from 0.5 to around 0.8. The
success rate is lowered then a little bit but reaches almost 0.9
after 20 generations and gets even higher after 25 generations. At
the same time, the median fee has been lowered down to a very
small value, around 1.0, which has been plotted in Fig. 3 (a). The
above two figures indicate that the learning process of the
players is ideal. We also plotted from the same simulation
evolutionary difference versus time in Fig. 4, which shows first
rapid, then slow and smooth decrease.\\
We next discuss the simulation results of our game where both
populations choose their bids on the basis of their friends' past
successful experiences in the same population. The friendship
strategy is implemented by constructing networks where players are
located. The four types of standard networks adopted in our
simulations are regular, small-world (as constructed by the
Strogatz-Watts method \cite{SW}), random (after Erd\"os-Renyi
\cite {ER}) and scale-free (as constructed by the Barabasi-Albert
algorithm \cite{BA}). In order to make the results of different
network topologies comparable, we set the average connectivity of
all networks to be fixed, for instance, 4. Here the offer of a
player at the present round is a function of the average of his
friends' last-round successful offers. In case there is no
successful friend in the previous round, he will simply select a
random value between 0 and $K$. One minor thing is that the
successful experiences of a player himself can also be considered.
But according to our simulations including it or not does not
cause too much difference. Fig. 3 (b) shows the success rate
versus the time $t$ when both populations are located on regular
friendship networks. Surprisingly, the improvement of the success
of deals here is even more remarkable than what has been shown in
Fig. 2 (a). We see that after only around 25 generations the
success rate approaches 1.0 and then remains quite stable.  Fig. 3
(b) gives the median fee versus the time $t$. Again, this
difference can be significantly decreased. We have also run
extensive simulations of our game when the network topologies are
small-world, random and scale-free. These simulations display
similar behaviors as what we have seen in Fig. 2 (b) and Fig. 3
(b). Another finding is that there is no network type among the
four ones considered that is systematically superior to the other
ones. Also, adding another layer of complexity, namely an
evolutionary optimization of networks based on the combined
fitnesses of their members does not lead to a particular network
paradigm.  Fig. 5 shows part of our simulations when buyers and
sellers have different friendship network strategies. The outcome
depends more on the (random) initial conditions than on the
network types involved.
\\We next investigate whether more available information will be
definitely advantageous to the players who have them. Buyers and
sellers will now be allowed to utilize different amounts of
information. More concretely, we say that they have a 1-round
opponent strategy when the members of the population can only look
one round back (they recall only their opponents' bids of the most
recent round). If they can look two rounds back, then they use a
2-round opponent strategy. Obviously, the players who have a
2-round opponent strategy have access to more information than
those who have the 1-round opponent strategy. We then compare the
intrinsic dynamic of a game with both populations using a 1-round
opponent strategy with the one where both use 2 rounds. The
comparison is plotted in Fig. 6. We note that when the members of
both populations can use only the most recent round information
the learning process is much quicker and more efficient than when
they have a 2-round opponent strategy. In the 1-round opponent
strategy simulation, the success rate becomes optimized and stable
after only 10 generations. While in the 2-round one, it takes
almost 100 generations to reach the same position as the former
one. To make the comparison more convincing we can play 1-round
opponent strategy against 2-round opponent strategy. This means we
let, for example, buyers take a 1-round opponent strategy and
sellers adopt a 2-round opponent strategy, or reversely. The
simulation of such a game will allow us to check whether the
symmetry will be broken, whether an equilibrium will be reached,
and most importantly, for which side the equilibrium value will be
more favorable. Our extensive simulations show that the
equilibrium will favor the side holding less information, i.e.,
with 1-round opponent strategy. Fig. 7 (a) shows one simulation
where buyers can recall one round back and the sellers can recall
two. As we can see, the equilibrium value of buyers is 20, smaller
than 25.  Fig . 7 (b) shows the reverse case of Fig. 7 (a) and the
situation then is more favorable to the sellers who have less
information.\\
It is also interesting to compare the value of the information
obtained from the successful experiences of friends and that from
one's own past experiences based on the opponents' behaviors. So
we can let one population choose the friendship network strategy
and another side have a 1-round opponent strategy or a 2-round
opponent strategy. Again, there are many possibilities here. Our
simulations indicate that in most cases, but not always, the
1-round opponent strategy gains an advantage over friendship
network strategies. As shown in both Fig. 8 (a) and Fig. 8 (b),
the equilibrium values here are more favorable to the side taking
1-round opponent strategy. Simulations were also done for other
friendship network topologies against the 1-round opponent
strategy and showed similar results. Another finding of our
simulations was that the play between the friendship network
strategies and the 2-round opponent strategy does not lead to a
decisive conclusion. In some runs, the population with the network
strategies can achieve a better position. But when the initial
conditions are changed, the situation may favor the population
with the 2-round opponent strategy. Two of our simulations of
friendship network strategy against 2-round opponent strategy were
plotted in Fig. 9. One interesting byproduct of our simulations is
that a population with simpler friendship network strategies, like
averaging over all friends without distinguishing them by success
or not, have better chances to win over the population with
1-round opponent strategy than that with more refined friendship
network strategies, like averaging (or even weighted, or evolved
weighted)  only over successful friends. This important finding
suggests again, among other things, that simpler and more flexible
strategies may be good choices  in the competition.\\
The following simulations support our conclusion that the achieved
equilibrium may favor the population with more restricted
strategic options. In such a case, each member of one population
takes the simplest strategy discussed above, of selecting a fixed
value as his offer. The simplest strategy is then subject to a
selection process with only random mutations occurring at a very
small probability. The members of the other side, however, can
have richer and more complex strategies. They could have a 1-round
opponent strategy or a network friendship strategy. We played the
two populations against each other and found that in most cases
the simple strategy can get a slight advantage over the 1-round
opponent strategy, shown in Fig. 10. The play of the simple
strategy against the friendship network strategies displays,
however, a more interesting behavior. As shown in Fig. 11 (a),
initially the population with the simple strategy has a prominent
advantage over the opponent side with the friendship network
strategies. We notice that the evolution of the population with
the friendship network strategies is almost suppressed and appears
rather slow. As the time goes on and the generations continue,
however, the population with the network strategies may have a
good chance to win over its opponent side, see Fig. 11 (b) and
(c). But such a process takes much longer time than ever required
in other simulations. For example, to acquire a better position,
the population may need 1000 or more generations.

\vskip 0.2cm
\begin{center}
\bf Figure Captions:
\end{center}

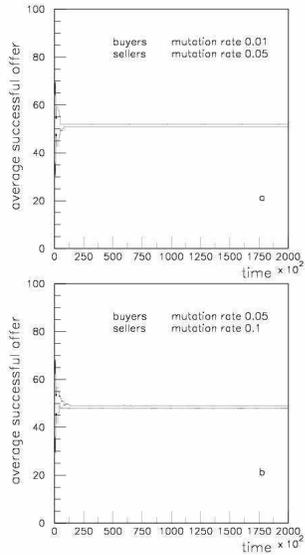
Fig. 1: The average successful offer versus time when both
populations take simple strategies but with different random
mutation rates. Average successful offer includes the average
offer of successful buyers and that of successful sellers. (a)
Mutation rate in the buyer population is 0.01 and for the sellers, 0.05;
(b) Mutation rate for buyer population is 0.05 and for sellers, 0.1.
Simulations were done with $N=400$ and $k=99$.

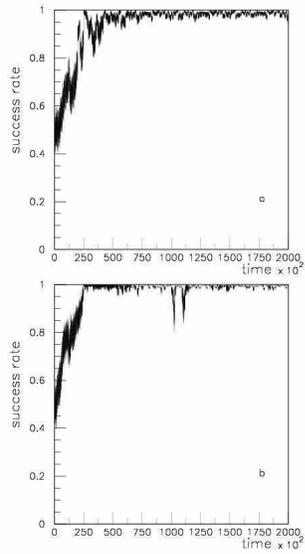
Fig. 2: Success rate versus time for the game where both
populations make offers according to (a) their own past
experiences (based on the behaviors of their opponents met in the
most recent round) and (b) their friends' successful experiences
(here players are interacted through regular networks). The
simulations were done with $N=400$ and $K=49$.

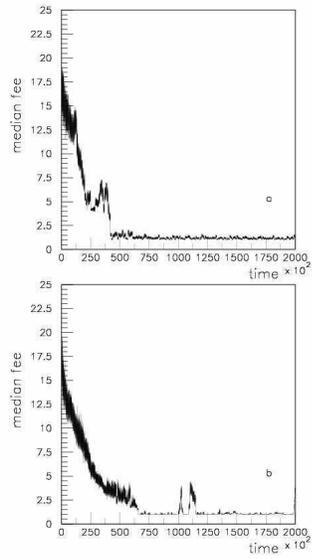
Fig. 3: Median fee versus time for the game where both populations
make offers according to (a) their own past experiences (based on
the behaviors of their opponents met in the most recent round) and
(b) their friends' successful experiences (here players are
interacting through regular networks). The simulations were done
with $N=400$ and $K=49$.

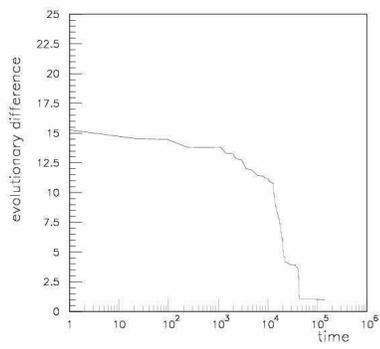
Fig. 4: Evolutionary difference versus time for the game where
both populations make offers according to their own past
experiences. The simulation was done with $N=400$ and $K=49$.

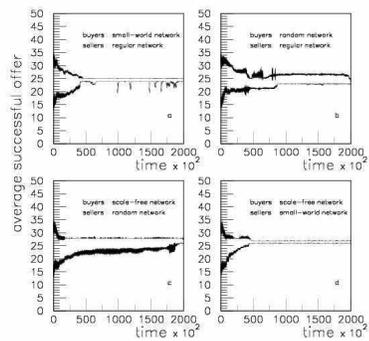
Fig. 5: The average successful offer versus time for friendship
network strategies playing against each other. (a) Buyers adopt a
small-world network strategy and sellers use a regular network
strategy; (b) Buyers take a random network strategy and sellers take
a regular network strategy; (c) Buyers take a scale-free network
strategy and sellers take a random network strategy; (d) Buyers take
a scale-free network strategy and sellers take a small-world network
strategy. Simulations were done with $N=400$ and $k=49$.

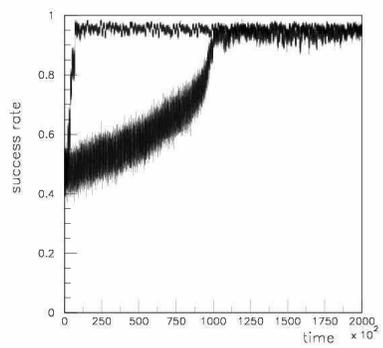
Fig. 6: Comparison of success rate versus time for two simulations
where both populations use the 1-round opponent strategy and the 2-round
opponent strategy, respectively. The top curve corresponds to the
game with 1-round opponent strategy and the bottom to two-round
opponent strategy. The simulations were done with $N=400$ and
$K=19$.

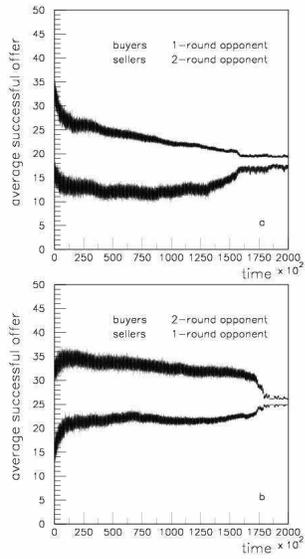
Fig. 7: The average successful offer versus time for 1-round
opponent strategy playing with 2-round opponent strategy. (a)
Buyers use the 1-round opponent strategy and sellers the 2-round
opponent strategy; (b) Buyers have the 2-round opponent strategy and
sellers have the 1-round opponent strategy. Simulations were done with
$N=400$ and $k=49$.

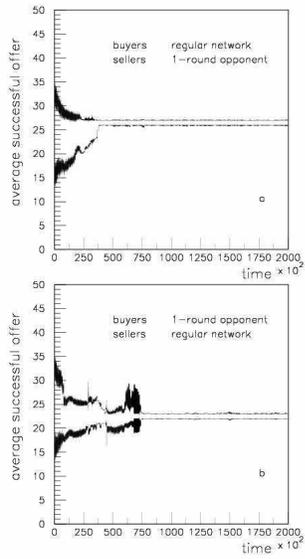
Fig. 8: The average successful offer versus time when friendship
network strategy plays against 1-round opponent strategy. (a)
Buyers have regular network strategy and sellers have the 1-round
opponent strategy. (b) Buyers have the 1-round opponent strategy and
sellers have the regular network strategy. Simulations were done with
$N=400$ and $k=49$.

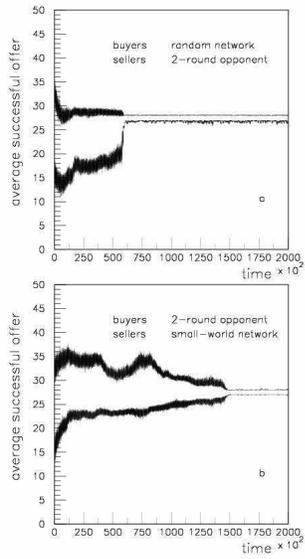
Fig. 9: The average successful offer versus time when friendship
network strategy plays against 2-round opponent strategy. (a)
Buyers take the random network strategy and sellers have the 2-round
opponent strategy. (b) Buyers use the 2-round opponent strategy and
sellers the small-world network strategy. Simulations were done
with $N=400$ and $k=49$.

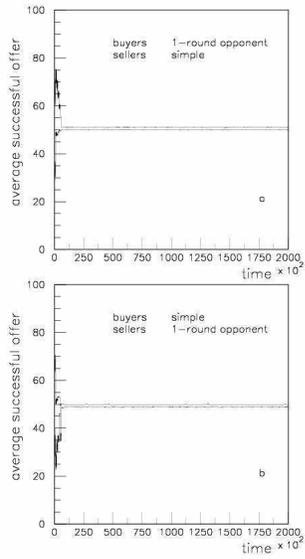
Fig. 10: The average successful offer versus time for simple
strategy playing against 1-round opponent strategy. (a) Buyers
adopt the 1-round opponent strategy and sellers the simple strategy;
(b) Buyers take the simple strategy and sellers have the 1-round opponent
strategy. Simulations were done with $N=400$ and $k=99$.

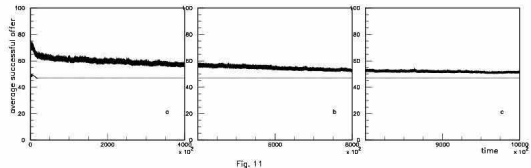
Fig. 11: Buyers use the simple strategy and sellers have the scale-free
network strategy. Here we show the average successful offer versus
time (a) from 0 to 400,000 time steps; (b) from 400,001 time steps
to 800,000 time steps and (c) from 800,001 time steps to 1000,000
time steps. Simulations were done with $N=400$ and $k=99$.

\end{document}